\documentclass[12pt]{article}
\usepackage{epsfig,graphicx}
\usepackage{amsmath}
\usepackage{amssymb}
\usepackage{array}
\usepackage{hyperref} 
\usepackage{units}
\newcommand{\ket}[1]{\:|\,{#1}\rangle\:}                      
\newcommand{\br}[1]{\:\langle{#1}|\:}

\newcommand{\lbr}{\bar{e}_R }

\newcommand{\Lbl}{\bar{L}_L}
\newcommand{\Qbl}{\bar{Q}_L}

\newcommand{\cO}{\mathcal{O}}
\newcommand{\cB}{\mathcal{B}}

\newcommand{\CL}{\mathcal{L}}

\newcommand{\vckm}{V}
\newcommand{\upmns}{U}

\newcommand{\LLFV}{{\Lambda_{\rm LFV}}}
\newcommand{\LLFVsq}{{\Lambda^2_{\rm LFV}}}
\newcommand{\LLFVq}{{\Lambda^4_{\rm LFV}}}
\newcommand{\LLN}{{\Lambda_{\rm LN}}}
\newcommand{\LLNsq}{{\Lambda^2_{\rm LN}}}

\newcommand{\hc}{\text{h.c.}}

\newcommand{\vu}{v_{u}}
\newcommand{\vusq}{v_{u}^2}
\newcommand{\vuq}{v_{u}^4}
\newcommand{\vd}{v_{d}}

\newcommand{\msq}{{m^2_\ell}}

\newcommand{\M}{{M_P}}
\newcommand{\Msq}{{M^2_P}}

\newcommand{\Gf}{{G_{\rm F}}}
\newcommand{\Gfsq}{{G^2_{\rm F}}}
\newcommand{\beq}{\begin{equation}}
\newcommand{\eeq}{\end{equation}}
\newcommand{\bea}{\begin{eqnarray}}
\newcommand{\eea}{\end{eqnarray}}
\newcommand{\no}{\nonumber}

\newcommand{\lsim}{
\mathrel{\hbox{\rlap{\hbox{\lower4pt\hbox{$\sim$}}}\hbox{$<$}}}}
\newcommand{\gsim}{
\mathrel{\hbox{\rlap{\hbox{\lower4pt\hbox{$\sim$}}}\hbox{$>$}}}}
\newcommand{\yuu}{\lambda_u}
\newcommand{\ydd}{\lambda_d}
\newcommand{\yee}{\lambda_e}

\newcommand{\yuuph}{\lambda_u^{\phantom{\dagger}}}

\newcommand{\yuuD}{\lambda_u^\dagger}

\newcommand{\yddT}{\lambda_d^T}

\newcommand{\ybu}{\bar\lambda_u}
\newcommand{\ybd}{\bar\lambda_d}
\newcommand{\ybe}{\bar\lambda_e}

\newcommand{\yfive}{\lambda_{5} }
\newcommand{\yten}{\lambda_{10} }
\newcommand{\ysig}{\lambda_{5}^\prime}

\newcommand{\ynu}{\lambda_{1}}

\newcommand{\ynuph}{\lambda_{1}^{\phantom{\dagger}}}

\newcommand{\LYGUT}{\CL_{\rm Y-GUT}}

\newcommand{\XQFV}{\Delta^{(q)}}

\newcommand{\DLL}{\Delta_{LL}}
\newcommand{\DRL}{\Delta_{RL}}
\newcommand{\nc}{\newcommand}
\nc{\ba}{\begin{array}}
\nc{\ea}{\end{array}}

\begin{document}

\begin{flushright}
FTUV/09-0616\\
IFIC/09-23
\end{flushright}
\vspace{1.0 true cm}
\begin{center}
{\Large {\bf  Violations of lepton-flavour universality in $P\to\ell\nu$ 
decays: a model-independent analysis} }\\
\vspace{1.0 true cm}
{\scshape Alberto Filipuzzi${}^a$ and Gino Isidori${}^b$}\\
\vspace{0.5 true cm}
${}^a$ {\small \sl Departament de F\'{i}sica Te\`{o}rica and IFIC, Universitat de Val\`{e}ncia-CSIC, \\Apt. Correus 22085, E-46071 Val\`{e}ncia, Spain} \\
\vspace{0.2 true cm}
${}^b$ {\small \sl INFN, Laboratori Nazionali di Frascati, Via E. Fermi 40, I-00044 Frascati, Italy} \\
\end{center}
\vspace{0.5cm}

\begin{abstract}
We analyse the violations of lepton-flavour universality in the 
ratios $\mathcal{B}(P\to\ell\nu)/\mathcal{B}(P\to\ell^\prime\nu)$
using a general effective theory approach, discussing various
flavour-symmetry breaking patterns of physics beyond the SM.
We find that in models with Minimal Lepton Flavour Violation the effects 
are too small to be observed in the next generations of experiments
in all relevant meson systems ($P=\pi,K,B$).
In a Grand Unified framework with a minimal breaking of the flavour 
symmetry, the effects remain 
small in $\pi$ and $K$ decays while large violations of  lepton-flavour 
universality are possible in $B\to\ell\nu$ decays. 
\end{abstract}

\section{Introduction} 

In the last few years there has been a great experimental progress in quark 
and lepton flavour physics. On the quark side, several stringent tests of 
neutral-current flavour-changing (FCNC) processes have been performed finding 
no significant deviations from the Standard Model (SM) predictions. This fact 
naturally points towards new physics models with highly constrained flavour structures. 
In models with flavoured degrees of freedom around the TeV scale, 
a natural option is the so-called hypothesis of Minimal Flavour Violation (MFV). 
Within a general Effective Field Theory (EFT) approach, this hypothesis 
can be formulated in general terms as the assumption that the Yukawa 
couplings are the only relevant sources of flavour symmetry breaking 
both within and beyond the SM, 
at least in the quark sector~\cite{D'Ambrosio:2002ex}. As discussed in the recent literature, 
this hypothesis can naturally be implemented in supersymmetric extensions 
of the SM~\cite{Paradisi:2008qh} or, in slightly modified forms,
also in models with new strongly interacting dynamics 
at the TeV scale~\cite{Cacciapaglia:2007fw,Kagan:2009bn}.

The situation of the lepton sector is more uncertain but also more exciting. 
The discovery of neutrino oscillations provide an unambiguous indication that 
the SM is not a complete theory and a clear evidence about the existence of new 
flavour structures in addition to the three SM Yukawa couplings. In various frameworks 
these new flavour structures can have non-trivial implications in other sectors of the 
model. In particular, deviations from SM predictions are expected in FCNC decays of 
charged leptons and, possibly, in a few meson decays with leptons pairs in the final
 state (see e.g. Ref.~\cite{Raidal:2008jk} for a recent review).

As far as meson decays are concerned, an interesting role is played by the helicity 
suppressed $P\to\ell\nu$ decays and particularly by the lepton-universality ratios 
in these processes. As pointed out in Ref.~\cite{Masiero:2005wr} 
(see also \cite{Ellis:2008st}), lepton-flavour violating 
effects in supersymmetric extensions of the SM could induce up to $\mathcal{O}(1\%)$ 
deviations from SM in the $R_K^{\mu e}=\mathcal{B}(K\to e\nu)\slash \mathcal{B}(K\to\mu\nu)$
ratio. Such level of precision is well within the reach of 
present~\cite{Antonelli:2008jg,Sciascia:2008kf}
and near-future experiments in kaon physics~\cite{Fantechi:2008zz}. A very significant test 
of lepton-flavour universality (LFU) will soon be performed also in $\pi\to\ell\nu$ decays, 
measuring the $R_\pi^{\mu e}$ ratio at the $\mathcal{O}(0.1\%)$ level \cite{Bryman:2007zz}. 
Finally, the purely leptonic decays of the $B^\pm$ mesons have just been observed at 
the $B$ factories~\cite{Ikado:2006un,Aubert:2007xj,Satoyama:2006xn,:2009rk}
opening the possibility of interesting LFU tests also 
in these modes~\cite{Isidori:2006pk}.

A general theoretical tool for analyzing the results of all these future experiments 
is provided by low-energy EFT approaches based on specific flavour-symmetry assumptions 
(such as the MFV hypothesis). This approach allows us to test the implications of flavour 
symmetries which are independent from specific dynamical details of the new-physics model. 
As far as lepton-flavour mixing is concerned, two general constructions of this type are 
particularly interesting: the EFT based on the Minimal Lepton Flavour Violation (MLFV) 
hypothesis~\cite{Cirigliano:2005ck} (see also~\cite{Cirigliano:2006su,Gavela:2009cd}) 
and the implementation of the MFV hypothesis in a 
Grand Unified Theory framework (MFV-GUT) \cite{Grinstein:2006cg}. The purpose of this 
paper is to analyze the role of the $P\to\ell\nu$ decays in both these
frameworks.

The paper is organized as follows: in Section~\ref{OpBasis} we analyze the 
dimension-six effective operators contributing to $P\to\ell\nu$ decays and 
discuss their flavour structure according to the MLFV hypothesis. 
In Section~\ref{DecayRates} we present the results for the $P\to\ell\nu$ 
decay rates and discuss their natural size in the MLFV framework.  
Section~\ref{MFV-GUT} is devoted the MFV-GUT framework and its peculiar 
features. The phenomenology is discussed in Section \ref{Phenomenology}, 
starting from the model-independent bounds derived from FCNC processes 
within the EFT approach. The final results concerning the maximal 
deviations from SM predictions on  $R_K^{\ell\ell^\prime}$ ratios are 
presented in Section~\ref{Phenomenology} and summarized in the Conclusions.

\section{Operator basis and flavour structures}
\label{OpBasis}

In order to construct the effective theory relevant 
to our analysis we introduce two energy scales: 
i)~the scale $\LLFV$, close to the electroweak scale, 
where new degrees of freedom carrying 
lepton-flavour quantum numbers appear; 
ii)~a very high energy scale $\LLN$ (or $M_\nu$),
well above $\LLFV$, 
associated to the breaking of total lepton 
number.
At low-energies we integrate out the new degrees 
of freedom and describe their effects by a series of 
non-renormalizable operators suppressed by inverse 
powers of $\LLFV$. 

In order to determine the relevant effective operators contributing 
to $P\to\ell\nu$ decays we need to specify the structure of the 
low-energy EFT. This is characterized by: i)~the low-energy field content; 
ii)~the flavour symmetry; iii)~the flavour-symmetry breaking terms. 
In all cases we assume that the low-energy field content of the EFT 
is the SM one, with the exception of the Higgs sector, where 
we we assume two Higgs doublets coupled separately to
up-type quarks ($H_u$), and down-type quarks and 
charged leptons ($H_d$). This way we avoid 
large contributions to FCNC, but we explore possible 
$\tan\beta = \langle H_u \rangle /\langle H_d \rangle$
enhancements of flavour-violating effects. 

As far as the flavour symmetry and symmetry-breaking terms are 
concerned, we consider first the two MLFV frameworks introduced 
in Ref.~\cite{Cirigliano:2005ck}, and briefly summarised below. 
Since in these two cases the deviations from the SM 
turn out to be very small, 
in Section~\ref{MFV-GUT} we analyse the GUT framework 
of Ref.~\cite{Grinstein:2006cg} with the aim 
of identifying if, and under which conditions, 
larger deviations from the SM are possible.

\subsubsection*{MLFV minimal field content}

In this case the lepton flavour symmetry group is
\beq
G_{\rm LF} = SU(3)_{L_L}\times SU(3)_{e_R}~, 
\eeq
and the lepton sector is invariant under two $U(1)$ symmetries 
which can be identified with the total lepton number, $U(1)_{\rm LN}$, 
and the weak hypercharge. In order to describe charged-lepton 
and neutrino masses we introduce the symmetry-breaking Lagrangian 
\begin{eqnarray}
\label{eq:Lag-SB_min}
\CL_{\text{Sym.Br.}}^{\rm min} &=& - \lambda_e^{ij} \,\bar e^i_R(H_d^\dagger L^j_L) -\frac1{2 \LLN}\,g_\nu^{ij}(\bar
L^{ci}_L\tau_2 H_u)(H_u^T\tau_2L^j_L)+\hc 
\end{eqnarray}
where $\lambda_e$ and $g_\nu$ are the two irreducible 
sources of $G_{\rm LF}$ breaking.\footnote{In Eq.~(\ref{eq:Lag-SB_min}) 
the indices $i,j$ are lepton-flavour indices (that in the following 
will often be omitted, or equivalently indicated up or down), 
and $\psi^c\equiv -i \gamma^2 \psi^*$. 
$L_L$ and $e_R$ denote the lepton doublet and the right-handed 
lepton singlet, respectively ($L_L^T \equiv (\nu_L,e_L)$,
$\bar{L}_L^c \equiv (\bar{\nu}_L^c, \bar{e}_L^c)$).
 We also denote by $\vu$ and $\vd$ the vacuum 
expectation values (vevs) of the Higgs doublets: $\vu\equiv v \sin\beta=\langle H_u \rangle$ 
and  $\vd\equiv v \cos\beta=\langle H_d \rangle$, where $v \approx 174$ GeV.} 
As anticipated, $\LLN$ denotes the scale of the flavour-independent
breaking of the $U(1)_{\rm LN}$ symmetry. 
The smallness of neutrino masses, $m_\nu \equiv g_\nu \vusq\slash \LLN$, 
is attributed to the smallness of $\vu / \LLN$.

It is convenient to treat the matrices $\lambda_e$ and $g_\nu$ 
as spurions of  $G_{\rm LF}$, such that the Lagrangian~(\ref{eq:Lag-SB_min}) 
and the complete  low-energy EFT is formally invariant under $G_{\rm LF}$.
The transformation properties of $\lambda_e$ and $g_\nu$ are 
\beq
\label{eq:spurion-rules}
\lambda_e \to V_R^{\phantom{\dagger}} \,\lambda_e V_L^\dagger~, \qquad\qquad
g_\nu \to V_L^{*\phantom{\dagger}} g_\nu V_L^\dagger~,
\eeq
where $V_L \in SU(3)_{L_L}$ and $V_R \in SU(3)_{e_R}$.
The simplest spurion combination controlling 
LFV transitions in the charged-lepton sector 
is $g_\nu^\dagger g^{\phantom{\dagger}}_\nu$.
Working in the basis where $\lambda_e$ is diagonal 
we can write
\beq
\label{eq:XLFV}
g_\nu^\dagger g^{\phantom{\dagger}}_\nu = \frac{\LLNsq}{\vuq}  ~ \upmns \bar{m}^2_\nu 
 \upmns^\dagger~,
\eeq
where $\bar{m}_\nu=$ diag($m_{\nu_1},m_{\nu_2},m_{\nu_3}$) 
and $\upmns$ is the usual PMNS mixing matrix 
(see Ref.~\cite{Cirigliano:2005ck} for notations).
Up to the overall normalization, the LFV spurion in 
(\ref{eq:XLFV}) is completely determined in terms of 
physical observables of the neutrino sector. 
Its explicit form in the case of  normal 
 hierarchy ($m_{\nu_1} < m_{\nu_2} \ll m_{\nu_3}$) or 
inverted hierarchy ($m_{\nu_3} \ll  m_{\nu_1} < m_{\nu_2}$) is
\bea
&& \left[ g_\nu^\dagger g^{\phantom{\dagger}}_\nu \right]_{ij}^{\rm(norm.)} = \frac{\Lambda_{\rm LN}^2}{v^4} \, \left[ 
m_{\nu_1}^{2}  \, \delta_{ij} \, + \, 
U^{\phantom{*}}_{i2} U_{j2}^* \, 
\Delta m^2_{\rm sol} 
+
U^{\phantom{*}}_{i3} U_{j3}^* \,  \Delta m^2_{\rm atm} \right] ~, \\
&& \left[ g_\nu^\dagger g^{\phantom{\dagger}}_\nu \right]_{ij}^{\rm(inv.)} = \frac{\Lambda_{\rm LN}^2}{v^4} \, \left[ 
m_{\nu_3}^{2}  \, \delta_{ij} \, + \, 
U^{\phantom{*}}_{i1} U_{j1}^* \, 
(\Delta m^2_{\rm atm}-\Delta m^2_{\rm sol}) 
+ 
U^{\phantom{*}}_{i2} U_{j2}^* \,  \Delta m^2_{\rm atm} \right] ~.
\eea
For simplicity, in the numerical analysis of the following sections 
we assume that the mass of the lightest neutrino vanishes 
($m_{\nu_1}=0$ in the normal hierarchy and $m_{\nu_3}=0$ in the inverted one).
Furthermore, we adopt the convention where $s_{13}\geq 0$ 
and $0\leq\delta< 2\pi$~\cite{PDG2008}.

\subsubsection*{MLFV extended field content}

In this case we assume the existence of three right-handed neutrino singlets 
under the SM gauge group, beside the SM degrees of freedom. The Majorana
mass matrix of these neutrinos is flavour-blind 
[$(M_R)_{ij}=M_\nu \: \delta_{ij}$], 
it is the only source of $U(1)_{\rm LN}$ breaking and it is assumed
to be much heavier that the electroweak scale 
($|M_\nu|\gg v$). The lepton flavour symmetry group is
\beq
SU(3)_{L_L}\times SU(3)_{e_R}\times O(3)_{\nu_R} = G_{\rm LF} \times O(3)_{\nu_R}~.
\eeq
The irreducible sources of flavour symmetry breaking are $\yee$ and $\lambda_\nu$, defined by
\beq \label{eq:LSymBreak_ext}
\CL_{\text{Sym.Br.}}^{\rm ext} = - \yee^{ij} \:\bar e^i_R(H_d^\dagger L^j_L) +i \lambda_\nu^{ij}\:\bar\nu_R^i(H_u^T \tau_2L^j_L)+\hc~,
\eeq
which have the following spurion transformation properties 
\beq
\yee \to V_R \yee V_L^\dagger \: , \qquad \lambda_\nu \to O_\nu \lambda_\nu V_L^\dagger \: ,
\eeq
with $V_L\in SU(3)_{L_L}$, $V_R\in SU(3)_{e_R}$ and $O_\nu\in O(3)_{\nu_R}$.
Integrating out the heavy right-handed neutrinos the effective left-handed Majorana 
mass matrix is $m_\nu = (v_u^2/M_\nu)\lambda^T_\nu\lambda^{\phantom{T}}_\nu$. 

In this framework the basic spurion combination 
controlling LFV transitions in the charged-lepton sector  
is $\lambda_\nu^\dagger \lambda^{\phantom{\dagger}}_\nu$. This can be unambiguously connected to 
the low-energy neutrino mass matrix only if we impose a further hypothesis, 
namely if we neglect CP violation in the neutrino mass matrix~\cite{Cirigliano:2005ck}:
\beq
\label{eq:XLFV2}
\lambda_\nu^\dagger \lambda^{\phantom{\dagger}}_\nu  ~\stackrel{\rm CP\ limit}{\longrightarrow}~
 \frac{M_\nu}{\vusq}  ~ \upmns  \bar{m}_\nu  \upmns^\dagger~.
\eeq
In the CP limit and neglecting the mass of the lightest neutrino,
we can write 
\bea
&& \left[ \lambda_\nu^\dagger \lambda^{\phantom{\dagger}}_\nu \right]_{ij}^{\rm(norm.)}
= \frac{M_\nu}{\vusq} \, \left[  
U_{i2} U_{j2} \, 
\sqrt{\Delta m^2_{\rm sol}} 
+ 
U_{i3} U_{j3} \,  \sqrt{\Delta m^2_{\rm atm}} \right]~, \\
&& \left[ \lambda_\nu^\dagger \lambda^{\phantom{\dagger}}_\nu \right]_{ij}^{\rm(inv.)}= \frac{M_\nu}{\vusq} \, \left[  
U_{i1} U_{j1} \, 
\sqrt{\Delta m^2_{\rm atm}-\Delta m^2_{\rm sol}} 
+ 
U_{i2} U_{j2} \,  \sqrt{\Delta m^2_{\rm atm}} \right] ~.
\eea

\subsection{Relevant effective operators and mixing matrices}

We are now ready to analyse the effective operator basis relevant 
for $P\to\ell\nu$ decays. We are interested in operators of dimension up to six,
invariant under the SM gauge group, 
that can contribute to LFV processes with a single charged-lepton-neutrino pair 
and a single meson. The presence of a single meson implies we can 
neglect all operators with a tensor Lorentz structure, because of their 
vanishing hadronic matrix element.  Moreover, we can safely neglect all the 
dimension-five operators, which necessarily 
break the total lepton number and are suppressed by
inverse powers of $\LLN$ or $M_\nu$.

The basic building blocks for the relevant dimension-six operators 
are the bilinears 
\beq
\bar{L}_L\DLL L_L \qquad \hbox{and} \qquad  \bar{e}_R\DRL L_L~,
\label{eq:Lbil}
\eeq
where $\DLL$ and $\DRL$ are spurions transforming 
under $G_{LF}$ as $(8,1)$ and ($\bar3$,3) 
respectively. The specific structure of the two spurions 
depends on the considered scenario. The lepton 
bilinears in~(\ref{eq:Lbil}) must be combined 
with corresponding quark bilinears, that we construct 
following the MFV rules~\cite{D'Ambrosio:2002ex}.
Restricting the attention to terms with at most one 
power of the quark Yukawa couplings,
the basis of relevant dimension-six operators is:
\begin{equation}
\label{eq:op_basis}
\begin{aligned} 
O_{RL}^{(1)}&=\lbr  \Delta_{RL} L_L ~\Qbl \ydd d_R \\
O_{RL}^{(2)} &= (D_\mu H_{d})^\dagger ~\lbr \Delta_{RL} D_\mu L_L\\
O_{RL}^{(3)} &= \lbr  \Delta_{RL} L_L^T ~ \bar{u}_R  \yuuD  i \tau_2 Q_L
\end{aligned} 
\qquad
\begin{aligned} 
O_{LL}^{(1)}&=\Lbl \gamma^\mu \tau_a \Delta_{LL} L_L~ \left( i D_\mu H_{u}\right)^\dagger
\tau_a  H_{u} \\
O_{LL}^{(2)}&= \frac{1}{2} ~ \Lbl \gamma^\mu \tau_a \Delta_{LL} L_L~ \Qbl \gamma_\mu \tau_a Q_L
\\
\\
\end{aligned} 
\end{equation}  
In principle an additional independent operator is obtained 
replacing  $H_u$ with $H_d$ in $O_{LL}^{(1)}$. However, this
operator has the same flavour and Lorentz structure as $O_{LL}^{(1)}$ 
and it is suppressed in the large $\tan\beta$ limit, therefore 
we ignore it.

Expanding the lepton spurions in powers of $\yee$, 
$g_\nu$ and $\lambda_\nu$ and retaining only the 
leading terms in the expansion (we assume all these 
couplings have perturbative elements), the
explicit form of $\DLL$ and $\DRL$ in the charged-lepton 
mass basis is:
\begin{itemize}
\item \textit{minimal case}\\
\beq
\DLL = \displaystyle \frac{\LLNsq}{\vuq} \upmns \: \bar{m}_\nu^2 \upmns^\dagger ~,
\qquad
\DRL = \ybe \left[\displaystyle \frac{\LLNsq}{\vuq} \upmns \: \bar{m}_\nu^2 \upmns^\dagger\right] ~,
\label{eq:DDLR1}
\eeq
\item \textit{extended case}\\
\beq
\DLL = \displaystyle \frac{M_\nu}{\vusq} \upmns \: \bar{m}_\nu \upmns^\dagger ~,
\qquad
\DRL = \ybe \left[\displaystyle \frac{M_\nu}{\vusq} \upmns \: \bar{m}_\nu \upmns^\dagger \right]~.
\label{eq:DDLR2}
\eeq
\end{itemize}

\section{$P\to\ell\nu$ matrix elements and decay rates}
\label{DecayRates}

Having defined the operator basis, we write the 
effective Lagrangian encoding new physics (NP) contributions as
\beq \label{L_eff}
\mathcal{L}^{\rm eff}_{\rm LFV} = \frac{1}{\LLFVsq}\sum_{n=1}^2 c_{LL}^{(n)} 
O_{LL}^{(n)} 
+\frac{1}{\LLFVsq} \sum_{n=1}^3 c_{RL}^{(n)} O_{RL}^{(n)} + \hc~,
\eeq
and compute the decay rates using  
$\mathcal{L}_{\rm tot}=\mathcal{L}_{\rm SM}+\mathcal{L}^{\rm eff}_{\rm LFV}$.
At the level of accuracy we are working, the SM amplitude 
for the $P^-\to\ell \bar \nu_\ell$ decay is 
\beq
A_{\rm SM}=\frac{4\Gf \vckm_{ab} }{\sqrt 2}\br{\ell \bar\nu_\ell }
\bar{e}_L^\ell \gamma^\mu \nu_L^\ell \ket{0} \br{0} 
\bar{u}_L^a \gamma_\mu d_L^b \ket{P^-}~.
\eeq
where $a,b$ are the quark-flavour indices of the corresponding meson
and $\vckm$ is the CKM matrix. Defining the meson decay constant,  
$\br{0}\bar{u}^{a}\gamma_{\mu}\gamma_{5}d^{b}\ket{P^{-}(p)}=i \sqrt{2} F_P p^{\mu}$,
the corresponding rate is
\beq
\Gamma(P^- \to\ell\bar\nu_\ell )_{\rm SM} = \frac{\Gfsq}{4\pi} 
\left|\vckm_{ab}\right|^2\: F^2_P \msq \M  
\left( 1-\frac{\msq}{\Msq} \right)^2~.
\eeq

The LL operators of $\mathcal{L}^{\rm eff}_{\rm LFV}$ 
have SM-like matrix elements, while the 
RL operators have a different structure:
\beq
A_{RL} \sim  \br{\ell \bar\nu_k}\bar{e}_R^\ell \nu_L^k \ket{0} \times \left\{
\begin{array}{l}
\br{0} \bar{u}^a \gamma_5  d^b \ket{P} \qquad\qquad O_{RL}^{(1)}~, O_{RL}^{(3)}~,\\
\\
(p_\ell)_\mu  
\br{0} \bar{u}_L^a  \gamma_\mu d_L^b \ket{P} \quad\:\: O_{RL}^{(2)}~.
\end{array}
\right. \label{eff_amplitud}
\eeq
For light mesons ($\pi$ and $K$), the hadronic matrix 
element of the pseudoscalar current 
\beq
\br{0}\bar{u}^a\gamma_{5}d^b \ket{P^{-}(p)}=-i \sqrt{2} F_P B_0~,
\qquad 
B_0 = M^2_P/(m_a+m_b)
\label{eq:B0}
\eeq
leads to a substantial enhancement of the first two terms 
in Eq.~(\ref{eff_amplitud}). Looking at the complete structure 
of the RL terms,
it is easy to realise that $O_{RL}^{(1)}$ is the potentially 
dominant one: $O_{RL}^{(2)}$ is suppressed by the extra 
charged-current interaction needed to mediate 
$P\to\ell\nu$ decays, while $O_{RL}^{(3)}$ is suppressed by the small value of 
the up-quark mass (appearing because of $\lambda_u$).
On the other hand, there is no clear difference 
among the two LL operators. Since we are interested in evaluating 
the maximal deviations from the SM, we concentrate the following 
phenomenological analysis on the three potentially dominant terms:
$O_{RL}^{(1)}$, $O_{LL}^{(1)}$ and $O_{LL}^{(2)}$.

In order to analyse the relative strength of SM and 
new-physics contributions, for each of these 
operators we define the ratio
\beq
R_{P\ell\nu}[O^{(n)}_{XX}]=\frac{\Gamma(P^-\to\ell \bar\nu_\ell)_{\rm SM}
+\delta\Gamma(P^-\to\ell \bar\nu_\ell)_{\rm int}
+\sum_{k\not=\ell}\Gamma(P^- \to\ell \bar\nu_k)_{\rm LFV}}{
\Gamma(P^-\to\ell\bar\nu_\ell)_{\rm SM}}~,
\eeq
where $\delta\Gamma_{\rm int}$ takes into account the lepton-flavour-conserving 
contributions generated by $\mathcal{L}^{\rm eff}_{\rm LFV}$ (including the 
interference with SM amplitude). 
The explicit expressions of this ratio 
for the three dominant operators are:
\bea 
R_{P\ell\nu}[O^{(1)}_{LL}] 
&=&
\displaystyle  \left|1 + \frac{\: 2 v_u^2 c_{LL}^{(1)}  }{\LLFVsq} ~
\Delta_{LL}^{\ell\ell} \right|^2+\sum_{k\neq \ell}  
~\frac{\: 4 \vuq | c_{LL}^{(1)} |^2}{\LLFVq}~ |\Delta_{LL}^{\ell k}|^2 ~, \nonumber  \\  
R_{P\ell\nu}[O^{(2)}_{LL}] &=&
\displaystyle  \left|1 - \frac{c_{LL}^{(2)}}{\sqrt{2}\Gf\LLFVsq}~ 
\Delta_{LL}^{\ell\ell} \right|^2+\sum_{k\neq \ell} 
 \frac{| c_{LL}^{(2)}|^2}{2\Gfsq \LLFVq}~ |\Delta_{LL}^{\ell k}|^2~,\label{eq:R_1RL} \\
R_{P\ell\nu}[O^{(1)}_{RL}] &=&
\displaystyle \left|1 - \frac{c_{RL}^{(1)}}{ 2\sqrt{2}\Gf \LLFVsq} ~ 
\Delta_{RL}^{\ell\ell} ~\frac{ m_{d} B_0}{v_d m_\ell}\right|^2 
+ \displaystyle \sum_{k\neq \ell}\frac{|c_{RL}^{(1)}|^2}{8\Gfsq \LLFVq}
~ |\Delta_{RL}^{\ell k}|^2 \left( \frac{ m_d B_0}{v_d m_\ell}\right)^2~, \nonumber 
\eea
where $m_d$ is the mass of the down-type quark inside the hadron
(the apparent dependence from light-quark masses is canceled by the 
corresponding dependence of $B_0$).

A closer look to these expressions allows us to 
identify the potentially dominant terms and 
the maximal size of the NP effects. 
The basic hypothesis of our approach is that $\LLFV$ 
is around the TeV scale and that all the Wilson 
coefficients are at most of $\mathcal{O}(1)$. This implies 
that the dimensionless coefficients of the $\Delta_{LL}$ 
terms in Eq.~(\ref{eq:R_1RL}) are at most of $\cO(10^{-1})$.
The size of $\Delta_{LL}$ is controlled by the scale of 
lepton-number violation ($\LLN$ or $M_\nu$): as shown 
in~\cite{Cirigliano:2005ck}, within this general 
EFT approach the scale of lepton-number violation
cannot be too large because of the bounds from 
$\mu\to e \gamma$. We shall come back on the 
precise bounds from LFV processes in 
Section~\ref{Bounds}, 
here we simply note that an order 
of magnitude estimate give $\Delta_{LL} \sim 
\LLNsq \Delta m^2_{\rm atm}\slash v_u^4\lsim \mathcal{O}(10^{-4})$. 
Thus for $R_{P\ell\nu}[O^{(1)}_{LL}]$ and
$R_{P\ell\nu}[O^{(1)}_{LL}]$ the non-standard effect is
necessarily small and the interference terms dominate.

The coefficients of the  $\Delta_{RL}$ terms in 
$R_{P\ell\nu}[O_{RL}^{(1)}]$ are apparently 
enhanced by an inverse dependence from 
the lepton mass. However, as shown in 
Eqs.~(\ref{eq:DDLR1})--(\ref{eq:DDLR2}), in the MLFV framework
the $\Delta_{RL}$ spurion contains a charged lepton Yukawa 
that cancels this dependence: 
\beq 
\Delta_{RL}^{\ell k} \frac{ m_{d} B_0}{v_d m_\ell} \approx
(\tan\beta)^2 \frac{m_P^2}{v_u^2}
\left(\frac{\Delta_{RL}^{\ell k}}{\bar\lambda_e}\right)
\approx
(\tan\beta)^2 \frac{m_P^2}{v_u^2} \Delta_{LL}^{\ell k}~.
\label{eq:massLR}
\eeq
The above result implies that the non-standard effect 
in $R_{P\ell\nu}[O_{RL}^{(1)}]$ are: i) negligible for 
$\pi$ and $K$ decays, even for large $\tan\beta$;
ii) of similar size of those in $R_{P\ell\nu}[O_{LL}^{(1,2)}]$ 
in the $B$-meson case at large $\tan\beta$.
We thus conclude that in the MLFV framework,
both within the minimal and the extended 
field content, there is no way to generate large 
deviations from the SM in $P\to\ell\nu$ decays.

The key difference with respect to the case discussed 
in Ref.~\cite{Masiero:2005wr} is that the MLFV hypothesis imply 
the same helicity suppression for SM and non-standard amplitudes. 
A general EFT framework where such condition is not 
necessarily enforced is the MFV-GUT framework 
that we analyse below.

\section{Beyond the minimal case: MFV-GUT framework} 
\label{MFV-GUT}

Following Ref.~\cite{Grinstein:2006cg}, we consider 
the implementation of the MFV principle in a Grand Unified 
Theory (GUT) based on the $SU(5)$ gauge group: the 
three generations of SM fermions fall into a $\bar 5$
$[\psi_i\equiv(d^c_{iR},L_{iL})]$ and a 10 
$[\chi_i\equiv(Q_{iL},u^c_{iR},e_{iR}^c)]$ of $SU(5)$, 
and we add three singlets for the right-handed 
neutrinos $[N_i\equiv\nu_{iR}]$. The maximal 
flavour group is then reduced 
to $SU(3)_{\bar 5} \times SU(3)_{10}\times SU(3)_1$.

Introducing three Higgs fields, $H_5$, $H_{\bar 5}$ and $\Sigma_{24}$, 
with appropriate $U(1)$ charges to avoid tree-level FCNCs, 
the Yukawa Lagrangian defining 
the irreducible sources of flavour-symmetry breaking is 
\bea 
\LYGUT  &=&   \yfive^{ij} ~ \psi^T_i \chi_j  H_{\bar 5}  + 
             \yten^{ij} ~ \chi^T_i  \chi_j  H_{5}  + 
\frac{1}{M} (\ysig)^{ij} ~ \psi^T_i \Sigma_{24} \chi_j H_{\bar 5} \no \\
&& + \displaystyle \ynu^{ij} ~ N^T_{i} \psi_j  H_{5}  + 
				  M_R^{ij} N^T_{i} N_{j} + \hc
\label{eq:LY_GUT}
\eea 
Imposing the invariance of $\LYGUT$ under the  
$SU(3)_{\bar 5} \times SU(3)_{10}\times SU(3)_1$ group implies 
\begin{align}
&\yfive^{(\prime)} \to  V_{\bar 5}^* ~\yfive^{(\prime)}~ V_{10}^\dagger~, \qquad 
&&\yten \to  V_{10}^* ~\yten~ V_{10}^\dagger~, \\
&\ynu   \to  V_{1}^{*\phantom{\dagger}}   ~\ynuph~  V_{\bar 5}^\dagger~ , \qquad  
&&M_R   \to   ~V_{1}^{*\phantom{\dagger}}  ~M_R^{\phantom{\dagger}}~ V_{1}^\dagger ~ .
\end{align}
with $V_{\bar 5}\in SU(3)_{\bar 5}$, $V_{10}\in SU(3)_{10}$ and $V_{1}\in SU(3)_{1}$. 

The non-renormalizable term in (\ref{eq:LY_GUT}) has been 
introduced to break the exact GUT relations between down-type 
quark and charged-lepton masses,
which are know to be violated in the case of the first two 
generations.\footnote{~The high-scale vev 
of $\Sigma_{24}$ breaks $SU(5)$ preserving $SU(2)_L\times U(1)_Y$,
$\langle \Sigma_{24} \rangle = M_{\rm GUT}{\rm diag}(1,1,1,-3/2,-3/2)$.
Moreover, we assume $M \gg M_{\rm GUT}$, such that this 
non-renormalizable term is negligible but for the first 
two generations. }
Expressing the 
low-energy Yukawa couplings in terms of the high-energy ones
we have
\beq
 \yuu = a_u \yten~, \quad \ydd = a_d \left(\yfive +  \ysig\right)~, \quad  
 \yee^T = a_e  \left(\yfive - \frac{3}{2} \ysig\right)~, \quad
\lambda_\nu = a_\nu \ynu~,
\label{eq:y_gut}
\eeq 
where $a_i=\cO(1)$ are appropriate renormalization-group
factors and we have redefined the spurion $\ysig$ incorporating 
a suppression factor $\sim \langle \Sigma_{24} \rangle/M$.
In the basis where the down-type quark Yukawa coupling is diagonal, 
the complete set of low-energies Yukawa couplings assume the form:
\bea
&& \ydd=\ybd~, \qquad  
\yuu= \vckm^T \, \ybu \, \vckm^{\phantom{T}}~, \qquad 
\yee =   C^{T}  \, \ybe  \, G^*  
\label{eq:new_mix_ma} ~, \no \\
&& \left[\lambda_\nu^\dagger \lambda^{\phantom{\dagger}}_\nu\right]_{\rm CP-limit}
~=~ \frac{M_\nu}{\vusq} G^T \upmns  \bar{m}_\nu  \upmns^\dagger G^*~,
\label{eq:GUT_yuka}
\eea
where, in analogy with the MLFV case with extended field content, we have 
assumed that $M_R$ is flavour blind [$(M_R)_{ij}=M_\nu \: \delta_{ij}$]. 

The two new mixing matrices $C$ and $G$ appearing in (\ref{eq:GUT_yuka}) 
control the diagonalization of $\yee$ in the basis 
where $\ydd$ is diagonal. In the spirit of minimising 
the unknown sources of flavour symmetry breaking, 
in the following we  work in the limit $C=G=I$. 
This assumption, which is justified in the limit 
where we can neglect the breaking of the flavour 
symmetry induced by $\ysig$ 
($C,G \to I$ in the limit  $\ysig\to 0$~\cite{Grinstein:2006cg}),
allows us to express all mixing effects in terms 
of the CKM and the PMNS matrices.

In the GUT framework the number of independent spurion combinations
contributing to LFV processes is much larger than in the MLFV case;
however, only few of them can give rise to a substantial 
parametrical difference. The potentially most interesting effect 
is obtained replacing the  $\DRL$ spurion of the MLFV case with 
$\DRL^{\rm GUT} = \yuuph\yuuD\yddT$. In the basis where the 
charged-lepton Yukawa is diagonal, this take the form
\bea
[\DRL^{\rm GUT}]_{\ell k} = \left[ C \Delta^{(q)}\ybd G^\dagger \right]_{\ell k}^* \quad 
\stackrel{\rm C=G=I}{\longrightarrow} \quad \left[\Delta^{(q)}\ybd \right]_{\ell k}^* ~ ,
\label{eq:DeltaGUT}
\eea
where
\beq 
\XQFV_{ij} \equiv  (\vckm^\dagger \,  \ybu^2  \, \vckm)_{ij} \approx
\frac{m^2_t}{\vu^2} (\vckm)^*_{3i} (\vckm)_{3j}~.
\label{eq:DFCNC_q}
\eeq
The key feature of $\DRL^{\rm GUT}$ in (\ref{eq:DeltaGUT}) is the presence of the 
suppressed down-type Yukawa coupling on the right, and not on the left,
as in Eqs.~(\ref{eq:DDLR1})--(\ref{eq:DDLR2}). This 
could allow to overcome the SM helicity suppression in LFV processes
with light charged leptons and neutrinos of the third generation.

\section{Phenomenology} 
\label{Phenomenology}

\subsection{Bounds from FCNC processes} 
\label{Bounds}

One of the advantage of the EFT approach is that we can derive 
model-independent bounds on the coefficients of the
effective Lagrangian in (\ref{L_eff}) from experiments. In particular,
we can extract some interesting bounds from FCNC transitions
of charged leptons which receive tree-level contributions 
from the operators in (\ref{L_eff}). 

At present the most stringent bound is obtained by the bounds on the 
$\mu-e$ conversion in nuclei and in particular by this result:
\beq
\cB_{\mu\to e}=\sigma(\mu^-\hbox{Au}\to e^-\hbox{Au}) \slash \sigma(\mu^-\hbox{Au}\to \hbox{capture})<7\times 10^{-13}~\cite{PDG2008} ~ .
\eeq
Starting from the Lagrangian (\ref{L_eff}), assuming $\tan\beta \gg 1$  
and using the notation of \cite{Kitano:2002mt}, we get:
\bea
\cB_{\mu\to e}&=&\frac{m_\mu^5 }{\Gamma_{\rm capt} \LLFVq}
\Bigg|\left[\left(1-4s_w^2\right)V^{(p)}-V^{(n)}\right] c^{(1)}_{LL} \Delta_{LL}^{\mu e}
+\left(-V^{(p)} + V^{(n)}\right)c^{(2)}_{LL} \Delta_{LL}^{\mu e} 
\nonumber \\
&& + \left(\tilde{g}_{RS}^{(p)}S^{(p)}+\tilde{g}_{RS}^{(n)}S^{(n)}\right) c^{(1)}_{RL}
\left[\Delta_{RL}^{\mu e} + (\Delta_{RL}^\dagger)^{\mu e}\right]
\Bigg|^2 ~ ,
\eea
where $\tilde{g}_{RS}^{(p)} = 4.3 \lambda_d + 2.5 \lambda_s$, 
$\tilde{g}_{RS}^{(n)} =5.1 \lambda_d + 2.5 \lambda_s$ 
and $V^{(p,n)}$, $S^{(p,n)}$ are dimensionless nucleus-dependent 
overlap integrals, whose numerical value can be found in~\cite{Kitano:2002mt}.

\begin{table}[t]
$$
\renewcommand{\arraystretch}{1.2}
\begin{array}{|c|c|c|c|c|}
\hline
& \raisebox{-10pt}[0pt][0pt]{\hbox{MLFV~minimal}}
& \multicolumn{2}{c|}{\hbox{MLFV}\ \hbox{extended}}  & \\
{\rm Coefficients}
& \raisebox{-10pt}[0pt][0pt]{$\LLN=10^{13}~{\rm GeV}$}  & 
  \multicolumn{2}{c|}{M_\nu =10^{12}~{\rm GeV}} & \hbox{MFV-GUT}\\ 
&  & \multicolumn{1}{c}{\hbox{Norm.~hier.}} &
\multicolumn{1}{c|}{\hbox{Inv.~hier.}} &  \\
\hline
\raisebox{0pt}[15pt][7pt]{$|c^{(1)}_{LL}|~ (1~{\rm TeV}\slash\LLFV)^2$} & 
< 1.9\times 10^{-1} & 
< 2.7\times 10^{-2} &
< 3.5\times 10^{-2} & 
-
\\
\raisebox{0pt}[7pt][7pt]{$|c^{(2)}_{LL}|~ (1~{\rm TeV}\slash\LLFV)^2$} &  
< 5.3\times 10^{-1} & 
< 3.8\times 10^{-2} &
< 5.0\times 10^{-2} & 
-
\\ 
\raisebox{0pt}[15pt][10pt]{$ |c^{(1)}_{LR}| \left(\frac{\tan\beta}{50}\right)^2 
\left(\frac{1~{\rm TeV}}{\LLFV}\right)^2$} & 
< 36    & 
< 4.9   &
< 6.6   & 
< 6.3 \\
\hline
\end{array}
$$
\caption{\label{table:exp_limit}\small{Bounds on the Wilson coefficients
of the LFV effective Lagrangian, from $\mu\to e$ conversion, in different 
flavour symmetry breaking frameworks. In the
MFV-GUT case we report the values in the $C,G\to I$ limit (see text), 
and only for the potentially dominant operator ($O_{RL}^{(1)}$). }}
\end{table}

Barring accidental cancellations among the contributions of different operators,
expressing the $\Delta$'s in terms of neutrino masses and mixing angles, according 
to Eqs.~(\ref{eq:DDLR1})--(\ref{eq:DDLR2}), we extract the bounds on the ratios 
of Wilson coefficients and effective scales reported in 
Table~\ref{table:exp_limit}.\footnote{~The numerical values of neutrino masses
and mixing angles used in the analysis are those reported  
in Ref.~\cite{Cirigliano:2005ck}.}
For simplicity, in the case of the two MLFV frameworks we show the bounds obtained
for reference values of $\LLN$ and $M_\nu$: for different values of these 
energy scales, the bounds can easily be rescaled 
according to Eqs.~(\ref{eq:DDLR1})--(\ref{eq:DDLR2}). 
This overall normalization problem does not appear in the GUT case: here 
we report the bound in the $C,G\to I$ limit and we analyse only the case 
of the potentially dominant operator  $O_{RL}^{(1)}$.

\subsection{Predictions for the lepton universality ratios} 
\label{R_universality}

Using the general expressions for the ratios in Eq.~(\ref{eq:R_1RL}) and the bounds 
on the Wilson coefficients reported in Table~\ref{table:exp_limit}, we are ready to 
derive predictions for the possible deviations from the SM in $P\to\ell\nu$ decays.
The most interesting observables are the lepton universality ratios,
$R_P^{\ell\ell^\prime} = \cB(P\to \ell \nu) \slash \cB(P\to\ell^\prime \nu)$,
whose values can be computed with high accuracy within the 
SM~\cite{Cirigliano:2007xi}. We parametrise possible 
deviations from the SM in the $R_P^{\ell\ell^\prime}$ as follows:
\beq 
\frac{R_P^{\ell\ell^\prime}|_{\rm exp}}{R_P^{\ell\ell^\prime}|_{\rm SM}} \equiv 1+\Delta r_P^{\ell\ell^\prime}~.
\eeq

\begin{table}[t]
$$
\renewcommand{\arraystretch}{1.2}
\begin{array}{|c|cccc|}
\hline
  \raisebox{-5pt}[0pt][0pt]{\hbox{Operators}}
& \raisebox{-5pt}[0pt][0pt]{\hbox{MLFV~minimal}}
& \multicolumn{2}{c}{\hbox{MLFV}\ \hbox{extended}}  & 
  \raisebox{-5pt}[0pt][0pt]{\hbox{MFV-GUT}}
\\  & & 
\multicolumn{1}{c}{\raisebox{5pt}[0pt][0pt]{\hbox{Norm.~hier.}}} &    
\multicolumn{1}{c}{\raisebox{5pt}[0pt][0pt]{\hbox{Inv.~hier. }}} &  \\
& 
\multicolumn{4}{c|}{\raisebox{0pt}[15pt][10pt]{
${\rm Bounds~on}~\vert \Delta r_\pi^{e \mu} \vert:$}}
\\ 
O^{(1)}_{LL} & 
< 2.7 \times 10^{-6} & 
< 2.3\times 10^{-6} &
< 3.3\times 10^{-6} & 
-
\\
O^{(2)}_{LL} & 
< 7.7 \times 10^{-6} & 
< 3.3\times 10^{-6} &
< 4.7\times 10^{-6} & 
-
\\

O^{(1)}_{RL} & 
< 2.1 \times 10^{-7} & 
< 1.7\times 10^{-7} &
< 2.5\times 10^{-7} & 
< 6.3\times 10^{-5}
\\
&
\multicolumn{4}{c|}{\raisebox{0pt}[20pt][10pt]{
${\rm Bounds~on}~\vert \Delta r_K^{e \mu} \vert:$}}
\\ 
O^{(1)}_{LL} & 
< 2.7 \times 10^{-6} & 
< 2.3\times 10^{-6} &
< 3.2\times 10^{-6} & 
-
\\
O^{(2)}_{LL} & 
< 7.7 \times 10^{-6} & 
< 3.3\times 10^{-6} &
< 4.7\times 10^{-6} & 
-
\\
O^{(1)}_{RL} & 
< 5.0 \times 10^{-6} & 
< 4.1\times 10^{-6} &
< 5.9\times 10^{-6} & 
< 5.0\times 10^{-2}
\\
\hline
\end{array}
$$
\caption{\label{table:r_pi}\small{Bounds on
 $|\Delta r_\pi^{e \mu}|$ and 
 $|\Delta r_K^{e \mu}|$ and 
in the various symmetry-breaking frameworks.}}
\end{table}

The maximal allowed values for $\Delta r_\pi^{e\mu}$
and  $\Delta r_K^{e\mu}$ are shown in Table~\ref{table:r_pi}.  
As can be seen, in the two MLFV frameworks these values are
far too small compared to the experimental precision expected 
in future experiments: few$\times 0.01\%$ in $R_\pi$ and few$\times0.1\%$ in $R_K$. 
Undetectable effects are found also in the 
$B\to\ell \nu$ system, within the two MLFV frameworks.

In the MFV-GUT case the situation is definitely 
more interesting. According to the last column of 
Table~\ref{table:r_pi}, in such case $\Delta r_{K}^{e \mu}$ 
might be within the reach of future experiments. 
However, this result must be taken with some care. 
The bounds in Table~\ref{table:r_pi} are obtained saturating the 
constraints on the Wilson coefficients from $\mu\to e$ conversion:
in the MFV-GUT framework these are not very stringent 
and are saturated only for unnatural large values of the 
Wilson coefficients or unnatural low values of the effective scale
$\Lambda_{\rm LFV}$. 
Imposing natural constraints on the parameters of the EFT, 
such as $|c^{(1)}_{LR}| < 1$, $\tan\beta < 50$, and $\Lambda_{\rm LFV} > 1$~TeV, 
the deviations from the SM in $R_K^{e\mu}$
 are at most around $0.1\%$, beyond the 
reach of future experiments. 

The reason why our bounds on $\Delta r_{K}^{e \mu}$ are stronger 
with respect to the maximal effects discussed in Ref.~\cite{Masiero:2005wr},
even in the  MFV-GUT framework, 
is the assumption of minimal breaking of the flavour symmetry. 
We have assumed that both $M_R$ and the new mixing matrices 
$C$ and $G$ are flavour blind. In this limit the mixing structure of 
$[\DRL^{\rm GUT}]_{\ell k}$ is controlled by the CKM matrix. This implies 
that the enhancement of $[\DRL^{\rm GUT}]_{i 3}$ due to the 
large Yukawa coupling of the third generation is partially
compensated by the suppression of $|V_{i3}|\ll 1$  ($i=1,2$).
This suppression could be removed only  with 
new flavour-mixing structures. 
We thus conclude that if some violation of LFU will be observed 
in $\pi$ decays at the  $0.1\%$ level, or in $K$ decays at the $1\%$
level, this would unambiguously signal the presence of {\em non-minimal} 
sources of lepton-flavour symmetry breaking.

\begin{table}[t]
$$
\renewcommand{\arraystretch}{1.5}
\begin{array}{|c|c|c|c|}
\hline
& R_B^{\mu\tau} & R_B^{e\mu}   & R_B^{e\tau}  \\
\hline
{\rm SM} & 
\approx 4.4 \times 10^{-3} &
\approx 2.4 \times 10^{-5} & 
\approx 1.1 \times 10^{-7} 
\\
\hline
{\rm MFV-GUT~~(mod.~ind.)} & 
< 7.0\times 10^{-3} &
< 1.6\times 10^{-2} & 
< 7.4\times 10^{-5} 
\\
\hline
{\rm MFV-GUT}~~(|c^{(1)}_{LR}|< 1)&
< 6.5 \times 10^{-3} &
< 4.2\times 10^{-4} &
< 2.8\times 10^{-6} 
\\
\hline
\end{array}
$$
\caption{\label{table:B_case}\small{Bounds on the universality 
ratios $R_B^{\ell \ell^\prime}$ in the MFV-GUT framework.
The model-independent bounds  are obtained saturating the 
constraint on $c^{(1)}_{LR}$ reported in Table~\ref{table:exp_limit}
and imposing $\cB(B\to\tau\nu)=\cB(B\to\tau\nu)_{\rm SM}$ (see text).
The bounds in the last line are obtained evaluating all
the decays with the {\em natural} conditions
$|c^{(1)}_{LR}| < 1$, $\tan\beta < 50$ and $\Lambda_{\rm LFV} > 1$~TeV. }}
\end{table}

The only system were sizable violations of LFU universality 
can be produced in the minimal set-up we are considering 
is the case of $B\to \ell \nu$ decays (assuming the MFV-GUT framework). 
Here the large meson mass provides a substantial enhancement 
of the contribution of the $\Delta_{RL}$ terms (see Eq.~(\ref{eq:massLR})).
As a result, in the helicity suppressed modes 
$B\to \mu \nu$ and, especially, $B\to e \nu$,   
the contribution of the $\Delta_{RL}$ terms
is large enough to compete with the SM.

The enhancement of $\cB(B\to e \nu)$ allowed by the model-independent 
bound on $c^{(1)}_{LR}$ (Table~\ref{table:exp_limit}) is huge
($\sim 10^{3}\times$~SM), and even imposing 
$|c^{(1)}_{LR}| < 1$ large non-standard effects are possible.
The maximal deviations from the SM for the three LFU 
ratios\footnote{~The 
model-independent bound on $c^{(1)}_{LR}$ allow
a lepton-flavour conserving contribution 
to $B\to \tau \nu$, from $O^{(1)}_{RL}$, 
of the same order of the SM amplitude. 
Since the experimental measurement of $\cB(B\to\tau\nu)$
is consistent with the SM expectation and we have not systematically 
analysed all the lepton-flavour conserving 
dimension-six operators, in Table~\ref{table:B_case}
we report the model-independent bounds on the $R_B^{\ell \ell^\prime}$ 
assuming $\cB(B\to\tau\nu)=\cB(B\to\tau\nu)_{\rm SM}$.}
are shown in Table~\ref{table:B_case}.
Given the suppression of the  $B\to \ell \nu$ rates, 
the experimental sensitivity needed to go below these bounds
is still far from the presently available one.
However, it could possibly be reached in future with 
high-statistics dedicated experiments.

\section{Conclusions}

Within the SM $P\to \ell\nu$ decays are 
mediated only by the Yukawa interaction
(the decay amplitudes 
can indeed  be computed to an excellent accuracy in the gauge-less 
limit of the model). This implies a strong helicity suppression
and a corresponding enhanced sensitivity to possible physics beyond the SM.
In particular,  the lepton-flavour universality 
ratios $R_P^{\ell\ell^\prime}=\mathcal{B}(P\to\ell\nu)/\mathcal{B}(P\to\ell^\prime\nu)$,
which can be predicted within high accuracy,
are interesting probes of the underlying lepton-flavour symmetry-breaking 
structure.

In this work we have analysed the deviations from the SM 
in the $R_P^{\ell\ell^\prime}$ ratios 
using a general effective theory approach, employing different 
ansatz about the flavour-symmetry breaking structures of physics 
beyond the SM. The main results can be summarised as follows:
\begin{itemize}
\item
In models with Minimal Lepton Flavour Violation,
both in the minimal and in the extended version (as defined in~\cite{Cirigliano:2005ck}), 
we find that the effects are too small to be observed in the next generations 
of experiments in all relevant meson systems ($P=\pi,K,B$).
This is because the MLFV hypothesis,  by construction, implies the same
helicity suppression of $P\to\ell\nu$ amplitudes as in the SM.
\item
In a Grand Unified framework with a minimal breaking 
of the flavour symmetry (as defined in Section~\ref{MFV-GUT})
the effects remain small in $\pi$ and $K$ decays, 
while large violations of lepton-flavour universality are possible 
in the $B$ system (see Table~\ref{table:B_case}). These are possible
mainly because of the enhancement of the 
flavour-violating $B\to e \nu_\tau$ rate.
\end{itemize}

\paragraph{Acknowledgments}
The work of A.F.~is supported by an FPU Grant (MICINN, Spain). This work has
been supported in part by the EU RTN network
FLAVIAnet [Contract No.~MRTN-CT-2006-035482], by
MICINN, Spain [Grant No.~FPA2007-60323 (A.F.)] and by Consolider-Ingenio
2010 Programme No.~CSD2007-00042-CPAN.

\end{document}